%
\input harvmac  

\def\fonttest{y}

\ifx\boringfonts\fonttest\def\ninepoint{}
\else
\font\ninerm=cmr9\font\ninei=cmmi9\font\nineit=cmti9\font\ninesy=cmsy9
\font\ninebf=cmbx9\font\ninesl=cmsl9\font\ninett=cmtt9
\def\ninepoint{\def\rm{\fam0\ninerm}
\textfont0=\ninerm \scriptfont0=\sevenrm \scriptscriptfont0=\fiverm
\textfont1=\ninei  \scriptfont1=\seveni  \scriptscriptfont1=\fivei
\textfont2=\ninesy \scriptfont2=\sevensy \scriptscriptfont2=\fivesy
\textfont\itfam=\nineit \def\it{\fam\itfam\nineit} \def\sl{\fam\slfam\ninesl}
\textfont\bffam=\ninebf \def\bf{\fam\bffam\ninebf}
\def\tt{\fam\ttfam\ninett}\rm}
\fi

\hyphenation{anom-aly anom-alies coun-ter-term coun-ter-terms
dif-feo-mor-phism dif-fer-en-tial super-dif-fer-en-tial dif-fer-en-tials
super-dif-fer-en-tials reparam-etrize param-etrize reparam-etriza-tion}


%
%
%
\newwrite\tocfile\global\newcount\tocno\global\tocno=1
\ifx\bigans\answ \def\tocline#1{\hbox to 320pt{\hbox to 45pt{}#1}}
\else\def\tocline#1{\line{#1}}\fi
\def\toclead{\leaders\hbox to 1em{\hss.\hss}\hfill}
\def\tnewsec#1#2{\newsec{#2}\xdef #1{\the\secno}
\ifnum\tocno=1\immediate\openout\tocfile=toc.tmp\fi\global\advance\tocno
by1
{\let\the=0\edef\next{\write\tocfile{\medskip\tocline{\secsym\ #2\toclead\the
\count0}\smallskip}}\next}
}
\def\tnewsubsec#1#2{\subsec{#2}\xdef #1{\the\secno.\the\subsecno}
\ifnum\tocno=1\immediate\openout\tocfile=toc.tmp\fi\global\advance\tocno
by1
{\let\the=0\edef\next{\write\tocfile{\tocline{ \ \secsym\the\subsecno\
#2\toclead\the\count0}}}\next}
}
\def\tappendix#1#2#3{\xdef #1{#2.}\appendix{#2}{#3}
\ifnum\tocno=1\immediate\openout\tocfile=toc.tmp\fi\global\advance\tocno
by1
{\let\the=0\edef\next{\write\tocfile{\tocline{ \ #2.
#3\toclead\the\count0}}}\next}
}
%
%

%
%
%
%

\gdef\prlmode{F}
\long\def\optional#1{}
\def\cmp#1{#1}         
%
%
\let\narrowequiv=\equiv
\def\equiv{\;\narrowequiv\;}

\fontdimen16\tensy=2.7pt\fontdimen17\tensy=2.7pt 



%

%
%

\def\CT{{\cal T}}

%
%
%
\def\boxit#1#2{
        $$\vcenter{\vbox{\hrule\hbox{\vrule\kern3pt\vbox{\kern3pt
        \hbox to #1truein{\hsize=#1truein\vbox{#2}}\kern3pt}\kern3pt\vrule}
        \hrule}}$$
}




%




\def\splitexact#1#2{\mathrel{\mathop{\null{
\lower4pt\hbox{$\rightarrow$}\atop\raise4pt\hbox{$\leftarrow$}}}\limits
^{#1}_{#2}}}

%
%

%
%
%
%
\def\rmi{{\rm i}}

\def\ex#1{{\rm e}^{#1}}                 
\def\dd{\mskip 1.3mu{\rm d}\mskip .7mu} 



%
%

\def\IM{isomorphism}

%
%

\ifx\boringfonts\fonttest
\font\blackboard=cmssbx10 \font\blackboards=cmssbx10 at 7pt  
\font\blackboardss=cmssbx10 at 5pt
\else
\font\blackboard=msym10 \font\blackboards=msym7   
\font\blackboardss=msym5
\fi
\newfam\black
\textfont\black=\blackboard
\scriptfont\black=\blackboards
\scriptscriptfont\black=\blackboardss


%
\ifx\boringfonts\fonttest
\font\gothic=cmssbx10 \font\gothics=cmssbx10 at 7pt  
\font\gothicss=cmssbx10 at 5pt
\else
\font\gothic=eufm10 \font\gothics=eufm7
\font\gothicss=eufm5
\fi
\newfam\gothi
\textfont\gothi=\gothic
\scriptfont\gothi=\gothics
\scriptscriptfont\gothi=\gothicss

{\catcode`\@=11\gdef\oldcal{\fam\tw@}}
\newfam\curly
\ifx\boringfonts\fonttest\else
\font\curlyfont=eusm10 \font\curlyfonts=eusm7
\font\curlyfontss=eusm5
\textfont\curly=\curlyfont
\scriptfont\curly=\curlyfonts
\scriptscriptfont\curly=\curlyfontss
\def\cal{\fam\curly\relax}
\fi
%

\ifx\boringfonts\fonttest\else\fi

\global\newcount\pnfigno \global\pnfigno=1
\newwrite\ffile
\def\pfig#1#2{Fig.~\the\pnfigno\pnfig#1{#2}}
\def\pnfig#1#2{\xdef#1{Fig. \the\pnfigno}%
\ifnum\pnfigno=1\immediate\openout\ffile=figs.tmp\fi%
\immediate\write\ffile{\noexpand\item{\noexpand#1\ }#2}%
\global\advance\pnfigno by1}

%
%
\def\tfig#1{Fig.~\the\pnfigno\xdef#1{Fig.~\the\pnfigno}\global\advance\pnfigno
by1}

%
%
%
%
\def\figI{y}
\def\ifigure#1#2#3#4{
\midinsert
\ifx\figflag\figI
 \ifx\htflag\figI
 \vbox{
  \href{file:#3}
{Click here for enlarged figure.}}
 \fi
 \vbox to #4truein{
 \vfil\centerline{\epsfysize=#4truein\epsfbox{#3}}}
\else
\vbox to .2truein{}
\fi
\narrower\narrower{\ninepoint\noindent{\bf #1:} #2}
\endinsert
}








%
%

%


\def\inbar{\,\vrule height1.5ex width.4pt depth0pt}
\def\IB{\relax{\rm I\kern-.18em B}}
\def\IC{\relax\hbox{$\inbar\kern-.3em{\rm C}$}}
\def\ID{\relax{\rm I\kern-.18em D}}
\def\IE{\relax{\rm I\kern-.18em E}}
\def\IF{\relax{\rm I\kern-.18em F}}
\def\IG{\relax\hbox{$\inbar\kern-.3em{\rm G}$}}
\def\IH{\relax{\rm I\kern-.18em H}}
\def\II{\relax{\rm I\kern-.18em I}}
\def\IK{\relax{\rm I\kern-.18em K}}
\def\IL{\relax{\rm I\kern-.18em L}}
\def\IM{\relax{\rm I\kern-.18em M}}
\def\IN{\relax{\rm I\kern-.18em N}}
\def\IO{\relax\hbox{$\inbar\kern-.3em{\rm O}$}}
\def\IP{\relax{\rm I\kern-.18em P}}
\def\IQ{\relax\hbox{$\inbar\kern-.3em{\rm Q}$}}
\def\IR{\relax{\rm I\kern-.18em R}}
\font\cmss=cmss10 \font\cmsss=cmss10 at 10truept
\def\IZ{\relax\ifmmode\mathchoice
{\hbox{\cmss Z\kern-.4em Z}}{\hbox{\cmss Z\kern-.4em Z}}
{\lower.9pt\hbox{\cmsss Z\kern-.36em Z}}
{\lower1.2pt\hbox{\cmsss Z\kern-.36em Z}}\else{\cmss Z\kern-.4em Z}\fi}
\def\IGa{\relax\hbox{${\rm I}\kern-.18em\Gamma$}}
\def\IPi{\relax\hbox{${\rm I}\kern-.18em\Pi$}}
\def\ITh{\relax\hbox{$\inbar\kern-.3em\Theta$}}
\def\IOm{\relax\hbox{$\inbar\kern-3.00pt\Omega$}}


\def\cmp#1{#1 }         


\long\def\suppress#1{}
\suppress{\def\boringfonts{y}  

\baselineskip=20pt
\def\ifigure#1#2#3#4{\nfig\dumfig{#2}}

} 

\def\ifigure#1#2#3#4{
\midinsert
\ifx\figflag\figI
 \ifx\htflag\figI
 \vbox{
  \href{file:#3}
{Click here for enlarged figure.}}
 \fi
 \vbox to #4truein{
 \vfil\centerline{\epsfysize=#4truein\epsfbox{#3}}}
\else
\vbox to .2truein{}
\fi
{\baselineskip8pt\narrower\narrower\ninepoint\noindent{\bf #1:} #2\par}
\endinsert
}

\def\optional#1{}
\def\testp{T}

\Title{\vbox{\hbox{UPR--766T}
}}{New Measurements of DNA Twist Elasticity}

\centerline{Philip Nelson}\smallskip
\centerline{Department of Physics and Astronomy, University of Pennsylvania}
\centerline{Philadelphia, PA 19104 USA}
\bigskip

\ifx\prlmode\testp
\fi
\ifx\answ\bigans \else\noblackbox\fi
\baselineskip20pt

The symmetries of the DNA double helix require a new term in its
linear response to stress: the coupling between twist and stretch.
Recent experiments with torsionally-constrained single molecules give
the first direct measurement of this new material parameter. We
extract its value from a recent experiment. We also present a very
simple microscopic theory predicting a value comparable to the one
observed. Finally we sketch the effect of constrained twist on
{\it entropic} elasticity of DNA arising from the connection between
Link, Twist, and Writhe.

\vskip.7truein\leftline{{\sl Running Title:} DNA Twist Elasticity}
\leftline{{\sl Keywords:} Entropic elasticity; Optical tweezers; DNA
conformation}
\leftline{\sl Talk presented at the Workshop on DNA Topology, Rutgers
University, April 1997}
\Date{8 July  1997}\noblackbox
\def\ctwist{C}
\def\cstretch{B}
\def\cts{D}

\let\epsilon=\varepsilon
\let\phi=\varphi

\def\const{{\rm const.}}

\def\tot{_{\rm tot}}

\def\lk{{\rm Lk}}

\def\wo{{\omega_0}}

\def\thp{{\bf t}_\perp}
\def\EH{{\bf E}}

\def\kbt{k_{\rm B}T}

\def\ft{\tilde f}
\def\const{{\rm const.}}

\def\wo{\omega_0}

\def\zh{{\bf z}}\def\eh{{\bf e}}

\hfuzz=3truept\baselineskip20pt
\newsec{Introduction}
The idea of studying the response of DNA to mechanical stress
is as old as the discovery of the double helix structure itself. While
many elements of DNA function require detailed
understanding of specific chemical bonds (for example the binding of
small ligands), still others are quite nonspecific and reflect
overall mechanical properties. Moreover, since
the helix repeat distance of $\ell_0\approx3.4\,$nm involves dozens of
atoms, it is reasonable to hope that this length-scale regime would be
long enough so that the cooperative response of many atoms would
justify the use of a continuum, classical theory, yet short enough
that the spatial structure of DNA matters. Since moreover various
important biological processes involve length
scales comparable to $\ell_0$ (notably the winding of DNA onto
histones), the details of this elasticity theory are
important for DNA function.

Recently, techniques of micromanipulation via optical tweezers and
magnetic beads have yielded reliable numerical values for the bend
stiffness from the
phenomenon of thermally-induced entropic elasticity (Smith et al.,
1992 Bustamante et al., 1994; Vologodskii, 1994; Marko et al., 1995), as
well as the direct measurement of another elastic constant, the
stretch modulus, by exploring the force range 10--50pN
(Cluzel et al., 1996; Smith et al., 1996; Wang et al.,
1997). Significantly, the relation
between bending stiffness,
stretch modulus, and the
diameter of DNA turned out to be roughly as predicted from the
classical theory of beam elasticity (Smith et al., 1996), supporting the
expectations mentioned above.

Still missing, however, has been any direct physical measurement of the elastic
constants reflecting the {\it chiral} ({\it i.e.} helical) character
of DNA. Recent experiments with torsionally constrained DNA have
permitted the determination of one such constant, the coupling between
twist and stretch (Strick et al., 1995; Marko, 1997; Kamien et al.,
1997).  This
coupling may be relevant for the binding of the protein RecA to DNA,
which
stretches and untwists the DNA (Stasiak et al., 1982). We will explain why
this term is needed,
extract its value from the experiment,
and compare it to a the prediction of a
simple microscopic model to see that its magnitude is
in line with the expectations of classical elasticity theory. Finally
we will briefly sketch how to understand another phenomenon visible in
the data, the effect of constrained link on entropic  elasticity
(J.D. Moroz and P. Nelson, in preparation).

\newsec{Experiment}
DNA differs from simpler polymers in that it can resist twisting, but
it is not easy to measure this effect directly due to the difficulty of
applying external torques to a single molecule.
The
first single-molecule stretching experiments constrained
only the locations of the two ends of the DNA strand.
The unique feature of the experiment of Strick {\it et al.}  was the
added ability
to constrain the {\it orientation} of each end of the molecule.

We will study Fig.~3 of (Strick et al., 1995). In this experiment, a constant
force of 8pN was applied to the molecule and the end-to-end length
$z\tot$ monitored as the terminal end was rotated through $\Delta\lk$
turns from its relaxed state (which has $\lk_0$ turns). In this way the
helix could be over- or undertwisted by as much as $\pm10$\%. Over
this range of imposed linkage $z\tot$ was found to be a linear function
of $\sigma$:
\eqn\eexpt{\epsilon=\const-0.15\sigma\, {\rm\  where\ \ }
\sigma\equiv\Delta\lk/\lk_0\ {\rm\  and\ \ }
\epsilon\equiv(z\tot/z_{{\rm tot,}0})-1\ .}
Thus $\sigma$ is the fractional excess link and $\epsilon$ is the
extension relative to the relaxed state.
Eqn.~\eexpt\ is the experimentally
observed twist-stretch coupling.

The existence of a linear term in \eexpt\ is direct evidence of the
chiral character of the molecule, and its sign is as expected on
geometrical grounds: untwisting the molecule tends to lengthen it.
Still geometry alone cannot explain this result. Consider the outer
sugar-phosphate backbones of the DNA. Suppose that the twist-stretch
phenomenon were due to the straightening of these helical backbones
while they
maintained constant length, 0.6~nm per  phosphate, and constant
distance 0.9~nm from the center of the molecule. Then since  each
basepair step is $h=0.34\,$nm high, the circumferential length per step
is $\ell_c=\sqrt{.6^2-.34^2}\,$nm. The
corresponding twist angle per step is given by
$\theta= (\ell_c/2)/.9{\rm nm}=32^\circ$, roughly as observed.
Supposing now an extension by $\Delta h/h=\epsilon$, we find an
untwisting by $\sigma=\delta\theta/\theta=\const-\epsilon/2.0$, quite
different from what is observed, eqn.~\eexpt. We must seek an
explanation of the experimental result not in terms of a geometrical
ball-and-stick model but in the context of an elastic response theory.

\newsec{Simple Model}
We will begin by neglecting bend fluctuations (see below). A straight
rod 
under tension and torque will stretch and twist. We
can describe it by the reduced elastic free energy
\eqn\ets{{F(\sigma,\epsilon)
\over \kbt L}
={\wo^2\over2}\left[
\ctwist\sigma^2+\cstretch\epsilon^2
+2\cts\epsilon\sigma\right]
 -f\epsilon\ .}
Here $C$ is the twist persistence length, $\cstretch\approx1100\,{\rm
pN}/\wo^2k_BT\approx 78\,$nm is the stretch modulus
(Wang et al., 1997), and $D$ is the desired twist-stretch coupling.
$L$ is the relaxed total length, $\wo=2\pi/\ell_0=1.85/$nm, and
the reduced force $\ft=8\,$pN/$k_BT\approx
1.95/$nm in the experiment under consideration. For a circular beam
made of isotropic material the cross-term
$\cts$ is absent, since twisting is odd under spatial
inversion while stretching  is even. For a helical beam, however, we
must expect to find this term.

We now minimize $F$ with respect to $\epsilon$ at fixed
force with an imposed constraint on the overtwist
$\sigma$ to find
\eqn\esolna{\epsilon=\epsilon_{\sigma=0}
-{(\cts/\cstretch)}\sigma\ .}
Comparing to \eexpt, we obtain the desired result: $\cts=12\,$nm.

\newsec{Bend Fluctuations}
We have discussed the term linear in overtwist $\sigma$ in \eexpt. For
the highest-force curve at 8~pN this is the dominant effect. At lower
forces, however, it is quickly overwhelmed by an effect {\it
symmetric} under $\sigma\to-\sigma$, which we have so far
neglected. This effect is due to the {\sl coupling between applied overtwist
and thermal bend fluctuations}. We now sketch a simple, though
imprecise, analysis of this effect. The full analysis is qualitatively
similar (J.D. Moroz and P. Nelson, in preparation).

Since in this section we want to study {\it nonchiral, low-force}
effects, we will revert to a model of a fixed-length cylindrical rod
with bend and twist elasticity. We will consider small deviations from
the unstressed state of the rod, which we take to run along the \zh\ axis.
Initially we paint a straight stripe
on the outside of the unstressed rod. To describe the deformed rod, we
find at each point a triad of unit vectors $\{\EH_i(s)\}$, where $\EH_1$ is the
tangent to the curve determined by the rod centerline, $\EH_2\perp \EH_1$
is the normal vector from the centerline to the stripe, $\EH_3=\EH_1\times
\EH_2$, and $s$ is arclength. Let $\eta\equiv\wo\sigma$ be the
imposed excess helix density, and define the convenient reference
frame
$$\eh_1(s)\equiv{\bf x}\cos(\eta s)+{\bf y}\sin(\eta s)\ ;\quad
\eh_2(s)\equiv-{\bf x}\sin(\eta s)+{\bf y}\cos(\eta s)\ ;\quad
\eh_3\equiv\zh\ .$$

We can now describe the deformed rod by three small variables: the
projection $\thp\equiv t_1\eh_1+t_2\eh_2$ of $\EH_3$ to the xy plane
and the angle $\phi$
between {\bf x} and the projection of $\EH_1$ to the xy plane. We
propose to expand the elastic energy to quadratic order in these and
thus find the thermal fluctuations in harmonic approximation.
In terms of $\thp,\phi$ we find
$$\eqalign{\EH_1&=(1-\half{t_1}^2-\half\phi^2)\eh_1+\phi\eh_2-(t_1+t_2\phi)\zh\cr
\EH_2&=-(\phi+t_1t_2)\eh_1+(1-\half{t_2}^2-\half\phi^2)\eh_2+(-t_2+t_1\phi)\zh\cr
\EH_3&=(1-\half{t_1}^2-\half{t_2}^2)\zh+t_1\eh_1+t_2\eh_2\ .\cr
}$$
We may now differentiate with respect to arc-length to get the
body-fixed angular velocities
$\Omega_1\equiv\EH_3\cdot\dot\EH_2=-\tau_2+\phi\tau_1$,
$\Omega_2\equiv\EH_1\cdot\dot\EH_3=\tau_1+\phi\tau_2$,
$\Omega_3\equiv\EH_2\cdot\dot\EH_1=\eta+\dot\phi-\half\eta({t_1}^2+{t_2}^2)+\dot
t_1t_2$, where we abbreviated $\tau_1\equiv\dot t_1-\eta t_2$,
$\tau_2\equiv\dot t_2+\eta t_1$. Note that the formula for the Twist,
$\Omega_3$, is just a simple derivation of Fuller's formula (Fuller, 1978)
for the Writhe of a nearly-straight curve.

Our formulas become very compact if we introduce the complex variable
$\CT\equiv ({\bf x}+\rmi{\bf y})\cdot\thp$.  We then have
${t_1}^2+{t_2}^2=|\CT|^2$ and
${\tau_1}^2+{\tau_2}^2=|\dot\CT|^2$. Finally we expand in Fourier
modes: $\CT=\sum_q\alpha_q\ex{\rmi qs}$ and similarly with
$\phi$. Substituting into the elastic
energy $E/\kbt=\half\int\dd s\left[
A({\Omega_1}^2+{\Omega_2}^2)+C{\Omega_3}^2\right]$
yields the harmonic elastic energy
\eqn\eharm{E/\kbt=\half\sum_q\left[
Aq^2-C\eta q+\ft\right]\,|\alpha_q|^2+\half C\sum_q q^2|\phi_q|^2\ .}
We have introduced the applied stretching force $\ft\equiv f/\kbt$.

The physics of this formula is clear: twist fluctuations decouple from
bend fluctuations, but the imposition of nonzero {\it net} overtwist $\eta$
creates a new crossterm in ${\Omega_3}^2$, first-order in
$s$-derivatives, and this crossterm affects the long-scale {\it bend}
fluctuations. Indeed completing the square in \eharm\ shows that the
effect of $\eta$ is to {\it reduce the effective tension} from $f$ to
$f-{\kbt\over 4A} (C\eta)^2$. This effective reduction is what makes the
relative extension plummet as the overtwist $\eta\equiv\wo\sigma$ is increased
at
fixed $f$. This effect, combined with the intrinsic twist-stretch
effect from the previous section, explains qualitatively all the
phenomena in the  region of the experiment where linear elasticity is
valid. A more precise version of this calculation  also affords a
direct determination of the value of the twist stiffness $C$
(J.D. Moroz and P. Nelson, in preparation; C. Bouchiat and
M. M\'ezard, in preparation.). The high-force regime studied here is
free from some of the difficulties of the Monte Carlo approach
(Vologodskii et al., 1979; J.F
Marko and A.V. Vologodskii, submitted); in
particular, there is no need for any artificial short-length cutoff.

\newsec{Microscopic Model}
The elastic theory in \S3 was very general, but it
gave no indication of the expected magnitudes of the various
couplings.
To gain further confidence in our result, we have
{\it estimated} the expected twist-stretch coupling based on the measured
values of the other elastic constants and geometrical information
about DNA (Kamien, 1996; R.D. Kamien, T.C. Lubensky, P. Nelson,
and C.S. O'Hern, submitted). We used a simple, intuitive
microscopic picture of DNA
as a helical rod to show how twist-stretch coupling
can arise and get its general scaling with the geometric
parameters.  The model shows that the value of $\cts$ calculated above is
reasonable.

\newsec{Conclusion}
We have pointed out a strong twist-stretch coupling in
torsionally-constrained DNA stretching experiments, evaluated it,
argued that it reflects intrinsic elasticity of the DNA duplex, and
shown that the value we obtained is consistent with elementary
considerations from classical elasticity theory. We also showed how
the interplay between twist and writhe communicates a constraint on
link into the entropic elasticity of DNA, as seen in experiment.

\vskip.5truein
We would like to thank
D. Bensimon, S. Block, and J. Marko for their help and for
communicating their
results to us prior to publication, and W. Olson for discussions.
RK, TL, and CO were  supported in part by NSF grant DMR96--32598.
PN was supported in part by NSF grant
DMR95--07366.
\vfill\eject
\def\pagin#1{#1}
\def\cmp#1{#1}
{\parindent0pt\frenchspacing

Bustamante, C., J.F.~Marko, E.D.~Siggia and S.~Smith. 1994.
     \cmp{Entropic elasticity of lambda-phage DNA.}
  Science { 265}:1599\pagin{--600}.

Cluzel, P. , A. Lebrun, C. Heller, R. Lavery,
J.-L. Viovy, D. Chatenay, and F. Caron. 1996. \cmp{DNA: an extensible
molecule.} Science {271}:792\pagin{--794}

Fuller, F.B. 1978. Decomposition of the linking number of a
closed ribbon. Proc. Natl. Acad. Sci. USA {75}:3357\pagin{--3561}.

Kamien, R.D., T.C. Lubensky, P. Nelson,
and C.S. O'Hern. 1997. Direct Determination of DNA Twist-Stretch Coupling.
Europhys. Lett. {38}:237\pagin{--242}.

Marko, J.F.,
and E.D.~Siggia. 1995. Stretching DNA. Macromolecules {28}:%
8759\pagin{--8770}.

Marko, J.F. Stretching must twist DNA. 1997.
Europhys. Lett. { 38}:183--188.

Saenger, W.  1984.  Principles of nucleic acid structure.   Springer.

Smith, S.B., L.~Finzi and C.~Bustamante. 1992.
      \cmp{Direct mechanical measurements of the elasticity of
single DNA molecules       by using magnetic beads.}
    Science { 258}:1122\pagin{--6}.

Smith, S.B., Y. Cui, and
C. Bustamante. 1996. \cmp{Overstretching B-DNA: the elastic response of
individual double-stranded and single-stranded DNA molecules.}
Science {271}:795\pagin{--799}\optional{[12/15/95]}.

Stasiak, A. and E. Di\thinspace Capua. 1982. The helicity of DNA
in complexes with RecA protein. Nature {299}:185.

Strick, T.R., J.-F. Allemand, D. Bensimon, A. Bensimon, and V.
Croquette. 1996. \cmp{The elasticity of a single supercoiled DNA
molecule.}
Science {271}:1835\pagin{--1837}.

Vologodskii, A.V., V.V. Anshelevich, A.V. Lukashin, and
M.D. Frank-Kamenetskii. 1979. Statistical mechanics of supercoils and the
torsional stiffness of the DNA double helix. Nature {280}:%
294\pagin{--298}.

Vologodskii, A.V. DNA extension
under the action of an external force. 1994. Macromolecules {27}:%
5623\pagin{--5625}.

Wang, M.D., H. Yin, R. Landick, J. Gelles,
and S.M. Block. 1997.
\cmp{Stretching DNA with optical tweezers.}
Biophys. J. 72:1335--1346.


}

\bye